\documentclass[12pt]{iopart}

\usepackage{graphicx}%
\usepackage{multirow}%
\usepackage{amsmath,amssymb,amsfonts}%
\usepackage{placeins}
\usepackage{afterpage}
\usepackage{soul}
\setstcolor{red}
\usepackage[labelfont=bf]{caption}

\usepackage{url}
\usepackage{amsthm}%
\usepackage{mathrsfs}%
\usepackage[title]{appendix}%
\usepackage{xcolor}%
\usepackage{textcomp}%
\usepackage{manyfoot}%
\usepackage{booktabs}%
\usepackage{algorithm}%
\usepackage{algorithmicx}%
\usepackage{algpseudocode}%
\usepackage{listings}%
\usepackage{textcmds}%
\usepackage{lipsum}
\usepackage{tabularx}
\usepackage{hyperref}
\usepackage{svg}
\usepackage[numbers]{natbib}
\usepackage{graphicx}
\usepackage{caption}
\usepackage{fancyhdr}
\def\bx{{\bf x}}
\def\bh{{\bf h}}
\def\ba{{\bf a}}

\def\bc{{\bf c}}
\def\bd{{\bf d}}
\def\bm{{\bf m}}
\colorlet{mycolor}{red} 

\begin{document}

\title{\textbf{Scalable Boltzmann Generators for  equilibrium sampling of large-scale materials}} 

\author{Maximilian Schebek$^1$, Frank No\'e$^{1,2,3,4}$, Jutta Rogal$^{1,5}$}

\address{$^1$ Fachbereich Physik, Freie Universit\"at Berlin,
14195 Berlin, Germany}
\address{$^2$ Fachbereich Mathematik und Informatik, Freie Universit\"at Berlin,
14195 Berlin, Germany}
\address{$^3$ Microsoft Research AI for Science, 10178  Berlin, Germany}
\address{$^4$  Department of Chemistry, Rice University, Houston, Texas 77005, USA}
\address{$^5$ Initiative for Computational Catalysis, Flatiron Institute, New York, New York 10010, USA}
\ead{m.schebek@fu-berlin.de}

\vspace{10pt}
\begin{indented}
   \item[] \today
\end{indented}

\begin{abstract}
The use of generative models to sample equilibrium distributions of many-body systems, as first demonstrated by Boltzmann Generators, has attracted substantial interest due to their ability to produce unbiased and uncorrelated  samples in `one shot'. Despite their promise and impressive results across the natural sciences, scaling these models to large systems remains a major challenge.  In this work, we introduce a Boltzmann Generator architecture that addresses this scalability bottleneck with a focus on applications in materials science.  We leverage augmented coupling flows in combination with graph neural networks to base the generation process on local environmental information, while allowing for energy-based training and fast inference. Compared to previous architectures, our model trains significantly faster, requires far less computational resources, and achieves superior sampling efficiencies. Crucially, the architecture is transferable to larger system sizes, which allows for the efficient sampling of materials with simulation cells of unprecedented size. We demonstrate the potential of our approach by applying it to several materials systems, including Lennard-Jones crystals, ice phases of mW water, and the phase diagram of silicon, for system sizes well above one thousand atoms. 
The trained Boltzmann Generators produce highly accurate equilibrium ensembles for various crystal structures, as well as Helmholtz and Gibbs free energies across a range of system sizes, able to reach scales where finite-size effects become negligible.
\end{abstract}

\section{Introduction}\label{sec1}
Generative models are emerging as a promising alternative to traditional sampling techniques such as molecular dynamics (MD) and Monte Carlo (MC) simulations for generating equilibrium ensembles of many-particle systems. Unlike MD or MC, which explore the configuration space through sequential updates over time~\cite{tuckerman2023statistical, Frenkel2001-yl}, generative models aim to learn a direct mapping from a prior distribution, from which uncorrelated samples are readily available, to the equilibrium distribution of interest. 
The key advantage of such a model is that it  enables the generation of statistically independent equilibrium samples in `one shot', bypassing the long correlation times and slow mixing inherent to conventional methods. 
Consequently, generative models have the potential to significantly reduce the cost of molecular simulations, 
specifically when applied to the computation of thermodynamic ensemble averages and free energies.
Building on the seminal work by No{\'e} et al. on Boltzmann Generators~\cite{Noe2019}, the combination of generative models such as normalizing flows~\cite{papamakarios_flow, tabak_normflow} with statistical reweighting has been applied across diverse areas within molecular and condensed matter sciences. Examples include the direct generation of equilibrium configurations from pure noise and the transformation of simulated distributions to explore different potentials or thermodynamic conditions, with applications ranging from proteins~\cite{midgley2023flow, invernizzi_lrex,klein2023equivariant, plainer2025consistent} and small molecules~\cite{Rizzi2021, Rizzi2023,Olehnovics2024} to condensed phase systems such as liquids~\cite{coretti_learning, wirnsberger_lfep}, atomic solids~\cite{Wirnsberger_2022, ahmad_free_2022, Wirnsberger_2023,schebek2024efficient}, and molecular crystals~\cite{kohler2023rigid, Olehnovics2025}.

Despite their promising properties, scaling Boltzmann Generators to large system sizes remains a significant challenge due to high training costs and low sampling efficiencies. Consequently, most of the above examples have focused on proof-of-concept problems, such as Lennard-Jones clusters and small molecules with fewer than 100 atoms. In the context of materials science, the largest systems studied to date are atomic solids comprising approximately 500 particles. Yet, these models have demonstrated very low sampling efficiencies despite requiring nearly one GPU-year of training~\cite{Wirnsberger_2022}. As a result, their computational cost exceeds that of MD–based simulations by several orders of magnitude, rendering them impractical for applications to realistic systems. This limitation is particularly critical in condensed-phase systems, where far larger system sizes are crucial to ensure converged and physically meaningful predictions~\cite{Vega2008,partay2027poly,Polson2000}.  Consequently, recent work has focused on scaling Boltzmann Generators to larger systems by developing improved architectures capable of handling increased system sizes and on amortizing training costs through transferable models~\cite{tan2025scalable, klein2024transferable, schebek2024efficient}. Although initial progress has been made, extending Boltzmann Generators to systems with thousands of atoms remains beyond current capabilities.

In this work, we address this challenge by developing Boltzmann Generators that scale to large crystalline materials. In contrast to previous approaches targeting condensed-phase systems~\cite{Wirnsberger_2022, ahmad_free_2022, Olehnovics2025}, we adopt a locality assumption 
and base the generative process on local environments rather than the full configuration. This is achieved by  leveraging graph neural networks~\cite{gilmer2017neural} in combination with augmented coupling flows~\cite{Huang2020-vv}, which allows an energy-based training, not requiring any samples from the target distribution while being computationally highly efficient to evaluate at inference time.  We show that this local architecture not only allows to train models to far higher sampling efficiencies but also significantly reduces the training costs. A crucial feature of our architecture is its transferability across system sizes, thereby allowing models trained on small systems to be employed on much larger ones, which drastically reduces the computational cost compared to a direct training on large systems.
Conditioning the model on external parameters, such as thermodynamic states or the atomic interaction potential, further amortizes the training costs.  We demonstrate the potential of our approach by producing accurate equilibrium ensembles for 
a diverse set of systems, including crystalline phases of the Lennard-Jones potential as well as various parameterizations of the Stillinger-Weber potential~\cite{stillinger1985computer} ranging from monatomic water to germanium and silicon, and reporting accurate Gibbs and Helmholtz free energies for system sizes well above 1000 atoms.

\section{Boltzmann Generators}\label{sec:methods}
Within the canonical ($NVT$) ensemble, with fixed number of particles $N$, temperature $T$, and volume $V$, the equilibrium distribution is defined by the  Boltzmann distribution~\cite{tuckerman2023statistical, Frenkel2001-yl}
\begin{align}\label{eq:boltzmann}
    p(\bx) = \frac{e^{-u(\bx)}}{Z}\quad ,
\end{align}
where  \(u(\bx)=\beta U(\bx)\) is the reduced potential with the potential  energy \( U(\bx)\) of  configuration $\bx\in \mathbb{R}^{3N}$, the inverse temperature $\beta= 1/k_BT$, and Boltzmann's constant $k_B$. The normalization constant $Z = \int e^{-u(\bx)} d\bx$ is known as the configurational partition function, from which the reduced Helmholtz free energy \( f=\beta F \)  is defined as
\begin{equation}\label{eq:f_logz}
    f =  - \log Z \quad .
\end{equation}
 The free energy further includes a momentum contribution, which can be evaluated analytically~\cite{tuckerman2023statistical}. All absolute free energies reported in the current study include the momentum contribution.
 
Boltzmann Generators (BGs)~\cite{Noe2019} use a normalizing flow $f_\theta : \mathbb{R}^{3N} \rightarrow \mathbb{R}^{3N}$~\cite{papamakarios_flow,tabak_normflow} to transform samples $\mathbf{x}$ from a base distribution $q(\mathbf{x})$ into samples $\mathbf{x}' = f_\theta(\mathbf{x})$ of a distribution $q_\theta(\mathbf{x})$ that approximates $p(\mathbf{x})$. The base distribution is often chosen as a normal distribution, which allows independent and identically distributed (i.i.d.) samples to be drawn directly. The exact-likelihood property of normalizing flows allows to evaluate the generated distribution via the change-of-variable theorem  
\begin{equation}\label{eq:cov}
    q_\theta(\bx') = q(\bx) \,|\det J_{f_\theta}(\bx)|^{-1} \quad,
\end{equation}
such that the generated samples can be reweighted to the target distribution. The mapping $f_\theta$ is commonly implemented using continuous-time formulations such as neural ordinary differential equations (ODEs)~\cite{chen2018neural, lipman2022flow}, or in a discrete fashion based on coupling flows~\cite{dinh_nice_2014,dinh2017density}. While continuous flows offer smooth and expressive mappings by modelling transformations as solutions to ODEs, the evaluation of the Jacobian, $J_{f_\theta}$, often involves high computational costs, especially in high dimensions. Another drawback is the need for samples from the target distribution~\cite{lipman2022flow}, which are, in fact, the very quantities the method aims to generate.  Although methods for reducing the cost of Jacobian evaluation~\cite{Peng2025} and for continuous models that do not require target samples are active areas of research with some initial progress~\cite{havens2025adjoint,dern_2025_energy}, these approaches are not yet mature enough to scale to the system sizes relevant to this work.   Here, we therefore focus on coupling flows which provide efficient exact likelihood evaluations,  even though they may require deep architectures to achieve high expressiveness. 

In contrast to continuous flows, an important advantage of coupling flows is their compatibility with likelihood-based optimization schemes, allowing the flow to be trained requiring only knowledge of the potential energy function of the target distribution  but no i.i.d. samples. Specifically, the flow is trained by minimizing the Kullback-Leibler (KL) divergence between generated and target distribution,  $D_{\rm KL}(q_\theta||p)$. For the equilibrium distribution of the canonical ensemble (Eq.~\eqref{eq:boltzmann}), the corresponding loss function reduces to~\cite{Noe2019}
\begin{equation}\label{eq:single_loss}
    \mathcal{L}_{qp}(\theta)  = - \mathbb{E}_{\bx\sim q}\bigl[\log w(\bx)\bigr] \geq \Delta f_{qp}\quad,
\end{equation}
where $\Delta f_{qp}$ is the free energy difference between $q$ and $p$. The importance weights $w$ are defined using the reduced potentials of the base and target, $u_q$ and $u_p$, as
\begin{equation}
w(\bx) = e^{u_q(\bx) - u_p(f_\theta(\bx)) +\log|\det J_{f_\theta}(\bx)| } \quad,
\end{equation}
such that $w(\bx) \propto  p(f_\theta(\bx)) / q_\theta(f_\theta(\bx))$. Since $\Delta f_{qp}$ is typically unknown during training, $\mathcal{L}_{qp}$ offers only limited insight into convergence. An additional useful diagnostic is the Kish effective sample size~\cite{Kish1965-es}
%
\begin{equation}\label{eq:ess_kish}
    \text{ESS} = \frac{\bigl[\sum_i w(\bx_i)\bigr]^2}{\sum_i [w(\bx_i)]^2} \quad.
\end{equation}
The ESS provides an approximate measure of how many uncorrelated samples would yield a Monte Carlo estimator of comparable statistical quality, making it a problem-independent indicator of the flow model’s sampling performance.

In the context of targeted free energy perturbation (TFEP)~\cite{tfep}, the free energy difference between base and target, $\Delta f_{qp}$, can be directly obtained from the trained BG via 
\begin{equation}\label{eq:tfep}
\Delta f_{qp} = -\log \mathbb{E}_{\bx \sim q}\bigl[w(\bx)\bigr] \quad.
\end{equation}
BGs can thus offer significant improvements in computational efficiency over traditional free energy estimators, such as free energy perturbation (FEP)~\cite{zwanzig_fep} and multistate Bennett acceptance ratio (MBAR)~\cite{Shirts2008-mbar}, which require a chain of intermediate distributions between $q$ and $p$ to ensure sufficient overlap for convergence. In this context, BGs can  be seen as a practical implementation of TFEP~\cite{tfep}, analogous to the learned free energy perturbation approach introduced in Ref.~\cite{wirnsberger_lfep}.

\section{Scaling Boltzmann Generators to large systems}
\label{sec:scale_bg}
\begin{figure}
    \centering
    \includegraphics[width=\linewidth]{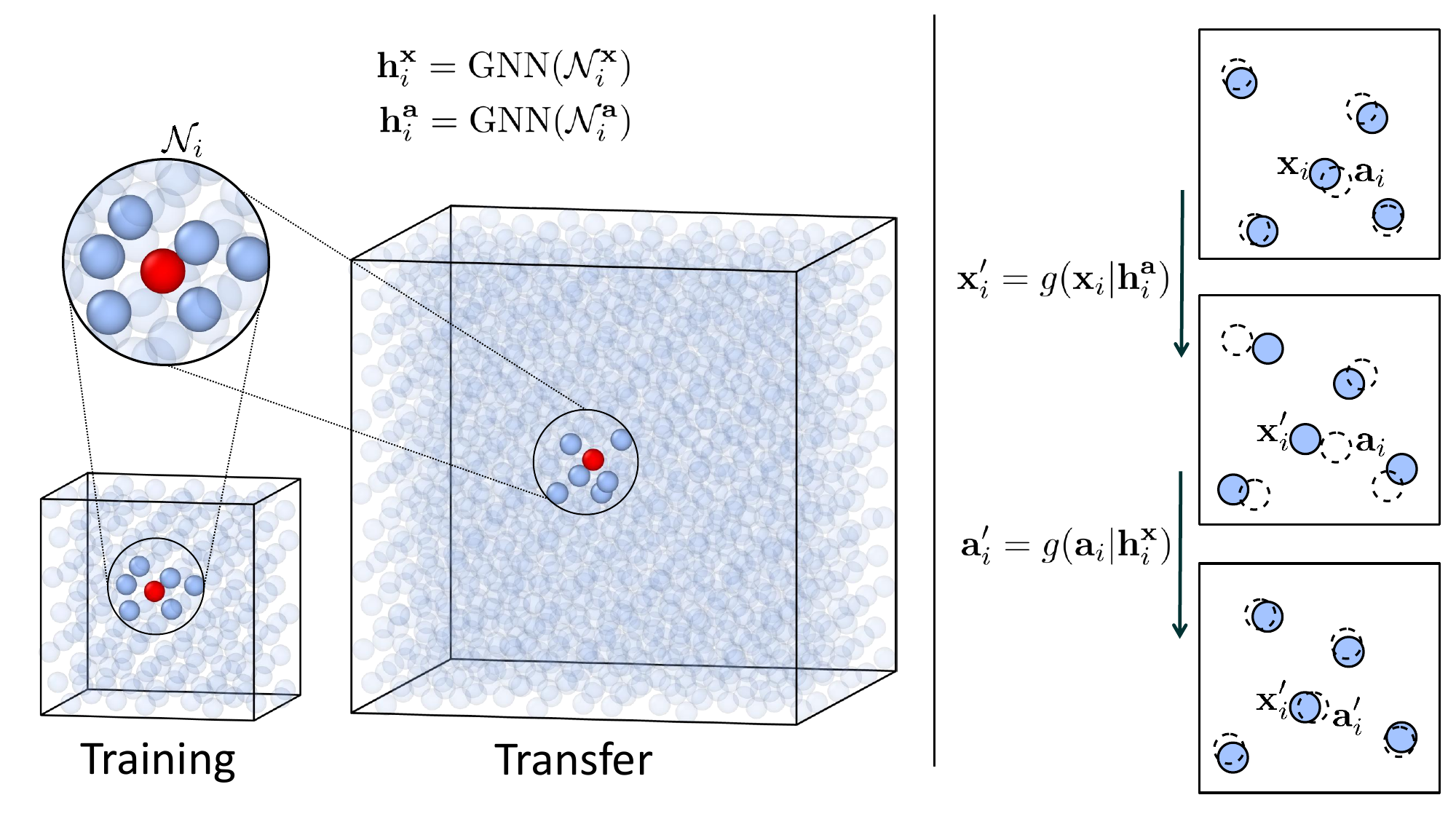}
    \caption{Scheme of the size-transferable augmented flow. Left: A GNN learns environment-dependent particle embeddings, $\mathbf{h}_i$, in a small system and is transferable across system size via each particle’s local neighborhood, $\mathcal{N}_i$. Right: Physical and auxiliary particles are updated sequentially using the learned embeddings. In each step, either all auxiliary particles or all physical variables are updated .   }
    \label{fig:scheme_method}
\end{figure}
Central to scaling any architecture in the context of many-body systems is formulating the learning problem based on local structural features~\cite{Behler2007Generalized,Musaelian2023}. 
Following this principle, we aim to train a flow to learn coordinate transformations based on local environments, allowing for efficient training on small systems and seamless transfer to larger ones. While in the context of continuous normalizing flows configurational updates can be defined based on the local, three-dimensional environments~\cite{satorras2021equivariant,klein2023equivariant}, this idea cannot be applied to flows based on coupling layers directly. Coupling layers require to split the input data into two channels and update one part of the data conditioned on the other. This procedure yields a triangular Jacobian whose determinant can be evaluated analytically, but the splitting either needs to be performed across particles or across spatial dimensions or a combination thereof, each of which prohibits the calculation of full three-dimensional environments~\cite{pmlr-v119-kohler20a}. Consequently, coupling-flow based architectures often rely on the absolute coordinates of all  atoms in the system~\cite{kohler2023rigid, Wirnsberger_2022}, and we will refer to this approach as a `global' architecture in the following.

The augmented coupling flow framework~\cite{Huang2020-vv} provides an elegant solution to overcome these architectural limitations by introducing auxiliary variables $\ba\in\mathbb{R}^{3N}$. Augmented flows enable the splitting to be performed between physical and auxiliary variables, thereby updating $\bx$ conditioned on $\ba$ and vice versa. Importantly, this scheme  retains full three-dimensional coordinates within the physical and auxiliary space. We encode spatial information into the auxiliary system by defining the joint base distribution of the physical and auxiliary system as
\begin{equation}\quad\label{eq:joint_base}
    q(\bx,\ba) =q(\bx)\,\mathcal{N}(\ba;\bx, \eta^2\mathbf{I}) \quad.
\end{equation}
Here, $\mathcal{N}(\cdot;\bx, \eta^2 \mathbf{I})$ denotes the normal distribution centered at $\bx$ with covariance matrix $\eta^2 \mathbf{I}$, where $\mathbf{I}$ is the identity matrix. The auxiliary system is therefore modeled as a noised copy of the physical system~\cite{midgley2023eacf}.  For a flow  acting on the augmented space, the change-of-variable for the augmented system holds similarly to Eq.~\eqref{eq:cov}:
\begin{equation}\label{eq:cov_augmented}
    q_\theta(\bx', \ba') = q(\bx,\ba) |\det J_{f_\theta}(\bx,\ba)|^{-1}\quad .
\end{equation}
The target distribution of the auxiliary variables is $\pi(\ba|\bx)=\mathcal{N}(\ba;\bx, \eta^2\mathbf{I})$, such that the flow is trained to optimize the joint target distribution
\begin{equation}
     p(\bx,\ba) = p(\bx)\, \mathcal{N}(\ba;\bx,  \eta^2\mathbf{I}) \quad.
\end{equation}
Similarly to the non-augmented case, training proceeds  by minimizing the KL divergence between  generated and  target distributions in the joint space. Within the augmented setting, only the joint generated density $q_\theta(\bx',\ba')$ can be evaluated exactly using Eq.~\eqref{eq:cov_augmented}. However, the marginal of the physical system can be obtained by integrating out the auxiliary degrees of freedom, as detailed in the  supplementary information (SI). Importantly, the evaluation of the marginal distribution only  requires an inverse pass through  the flow, but no evaluation of the target potential.

Utilizing the auxiliary variables, we define the coupling flow update for particle $i$  by
\begin{equation}
    \bx_i' = g(\bx_i|\bh^\ba_i)\quad, 
\end{equation}
where $g$ is a bijector parametrized by the embedding  $\bh^\ba_i$, which is computed using a graph neural network (GNN)~\cite{gilmer2017neural} that aggregates information from each particle’s local neighborhood (see Fig.~\ref{fig:scheme_method}). Details on the implementation can be found in the SI. A crucial feature of the local approach presented here is its linear scaling with system size, achieved by fixing the number of neighbors. This stands in stark contrast to global architectures based on, for example,  attention mechanisms which learn interactions between all particles, resulting in  quadratic scaling with system size~\cite{vaswani2017attention,Wirnsberger_2022}.

To ensure that the bijector $g$, which we implement as rational quadratic splines\cite{durkan2019neural}, remains applicable to larger system sizes, it is crucial that the magnitude of its input does not scale with the system size. One possibility that we follow in our proposed architecture is to model the displacements from the ideal crystal lattice rather than absolute positions. This makes the input size-independent and preserves the model’s ability to generalize to arbitrary system sizes. Accordingly, we define the physical base distribution as \(q(\bx)=\mathcal{N}(\bx;0,\eta^2_q\mathbf{I})\) and set $\eta=\eta_q$.

Within the augmented flow framework, the partition function of the generated joint distribution depends on both physical and auxiliary variables. Since the target and base distributions of the auxiliary system are Gaussians, the partition function factorizes (see SI) and the free energy of the augmented system is given by 
\begin{equation}
    f^{\rm aug} = f + f^{\rm aux} \quad ,
\end{equation}
where $f$  and $f^{\rm aux}$ denote the free energy of the physical (Eq.~\eqref{eq:f_logz}) and auxiliary system, respectively. 
The corresponding TFEP estimator (Eq.~\eqref{eq:tfep}) for the augmented system is 
\begin{align}   
\label{eq:tfep_aug}
    \Delta f^{\rm aug}_{qp} & = f_p^{\rm aug} - f_q^{\rm aug} =   f_p + f_p^{\rm aux} - (f_q + f_q^{\rm aux}) \\
        & = \Delta f_{qp} + \Delta f^{\rm aux}_{qp}\quad.
\end{align}
If $\eta$ of the auxiliary base and target distributions is chosen to be identical, then $\Delta f^{\rm aux}_{qp} = 0$ and $\Delta f_{qp}^{\rm aug} = \Delta f_{qp}$.

Flow-based approaches also allow to compute Gibbs free energies corresponding to the $NPT$ ensemble, in which the shape of the simulation box, $\bh\in \mathbb{R}^{3\times3}$ with $V=\det \bh$, changes during the simulation at constant pressure $P$~\cite{tuckerman2023statistical}. While prior work~\cite{Wirnsberger_2023, schebek2024efficient} studied flows modeling shape and configurational distributions simultaneously, 
in our current approach, the Gibbs free energy is computed via a Legendre transformation as $G = \min_\bh[F(\bh) + \det(\bh)  P]$~\cite{hashimoto2025efficient}. To obtain a reliable estimate for $F(\bh)$, we train the flow in a  shape-conditional fashion~\cite{schebek2024efficient,falkner2023conditioning}, optimizing the flow to generate samples for varying box dimensions (see SI for further details).

\section{Experiments}
The principle advantages of our proposed BG architecture based on local augmented flows are reduced training times, improved sampling efficiencies, and, most importantly, its ability to generalize to larger systems. We demonstrate the applicability of our method across a range of crystal structures for various materials systems,  
exemplified by the Lennard–Jones (LJ) potential and multiple parameterizations of the Stillinger–Weber  potential~\cite{stillinger1985computer}, including ice phases of monatomic water (mW) and silicon. Computational details are provided in the SI.

\subsection{Size transferable training and evaluation}

We illustrate the effectiveness of the size transferability by applying  local BGs, trained on cubic mW ice with \( N = 216 \) particles and face-centred cubic (FCC) LJ  with \( N = 256 \), to systems containing 512, 1000, and 1728 particles for cubic mW ice and 500, 864, 1000, and 1372 particles for FCC LJ.  

\begin{figure}[t]%
    \centering
    \includegraphics[width=\linewidth]{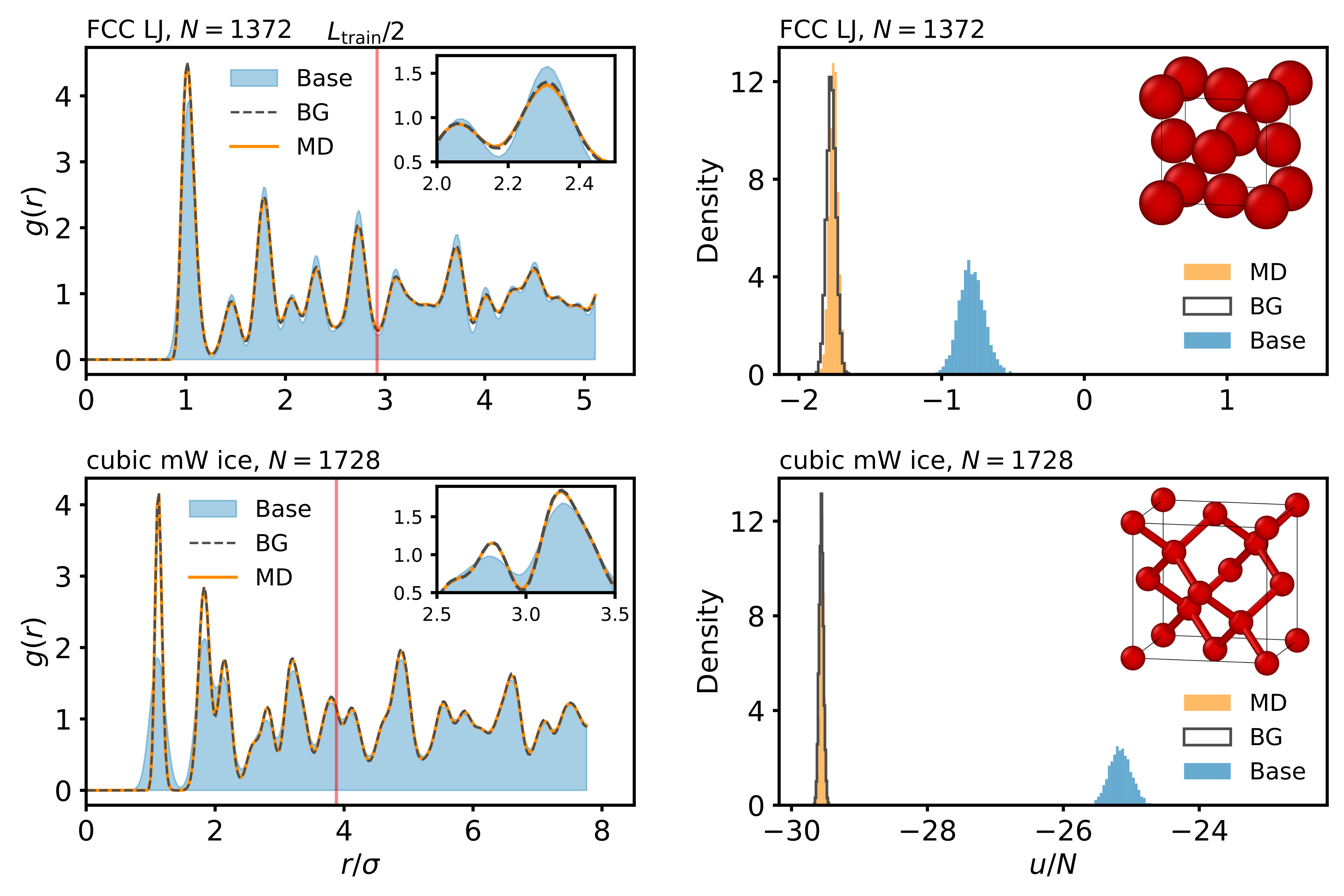}
  \caption{Radial distribution functions (left) in units of the interaction length $\sigma$ and reduced energy histograms (right)  for FCC LJ with  \( N = 1372 \) (top) and cubic mW ice with  \( N = 1728 \)    (bottom) as obtained from MD, the base distributions, and the local BGs. The local BG results were computed using models trained with \( N = 216 \) (mW) and \( N = 256 \) (LJ), with the red line indicating the corresponding half box length during training. No reweighting was applied. 
  \label{fig:rdf_transferable}}

\end{figure}

Figure~\ref{fig:rdf_transferable} shows the radial distribution functions (RDFs) and potential energy histograms for the largest investigated system sizes, as computed from samples generated by the BGs, MD, and drawn from the base distribution. For both systems, the RDFs produced by the BGs are nearly indistinguishable from those obtained via MD over length scales going far beyond the size of the training systems.
While by design the RDF of the base distribution already captures the main features of the target, the corresponding configurations do not reflect the correct correlations between particle positions in the target potential, which is evident from the pronounced differences in the  energy distributions of the base and MD ensembles. In contrast, the local BGs accurately reproduce both structural and energetic features in both systems, yielding excellent agreement in the histograms already without reweighting.  These results clearly demonstrate that coordinate transformations based on local environments are entirely sufficient for generating accurate equilibrium structures and are reliably transferable to larger system sizes.

While structural and energetic accuracy are necessary conditions for accurate free energy estimates, they alone do not guarantee high sampling efficiency or precise free energy calculations. A more rigorous measure is the effective sample size (ESS) (Eq.~\eqref{eq:ess_kish}) evaluated across the various system sizes within the transferable framework and summarized in the lower part of Table~\ref{tab:ess-results-transposed}. As the system size increases, the ESS naturally decreases, as even constant per-particle errors in the extensive potential energy lead to exponentially amplified errors in the Boltzmann distribution. But even for large systems such as cubic mW ice with \( N=1728 \) and FCC LJ with \( N=1372 \), the ESS of the local BGs is large enough to ensure reliable and highly accurate statistics of ensemble averages.

Importantly, the ESS for large systems is significantly higher than that obtained with global approaches, while the computational cost is significantly reduced. For mW ice with $N = 512$,  global BGs reported in Refs.~\cite{Wirnsberger_2022,Wirnsberger_2023} required more than 330 GPU days  of training  until convergence, yet achieved ESS values of only around 0.2\%. In contrast, for the mW system with $N = 216$, our local BG  reaches convergence in approximately four GPU days  on  the same hardware (see SI for details on training times), while yielding ESS values for the 512-particle system up to an order of magnitude higher than those obtained with the global BG. Moreover, scaling global architectures to larger systems becomes computationally prohibitive, making particle counts beyond 500 effectively impossible due to excessive training costs and deteriorating sampling efficiency.

\begin{figure}[t]%
    \centering
    \includegraphics[width=\linewidth]{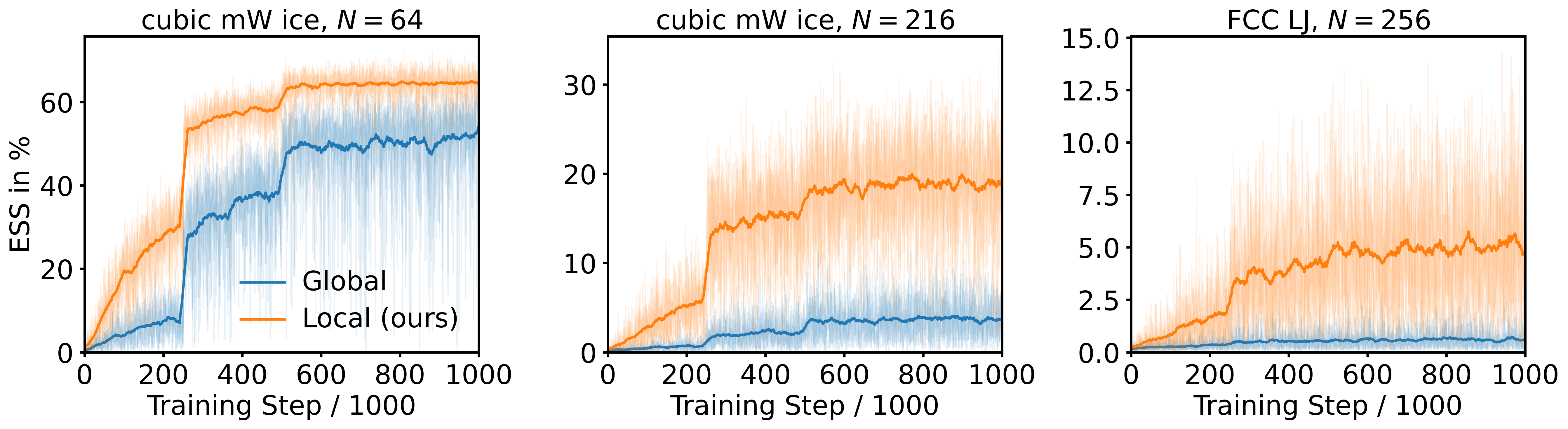}
    \caption{Effective sample size (ESS) plotted against the number of training steps for global and local  BGs of cubic mW ice with $N = 64$ and $N = 216$ (left and middle panels), and FCC LJ with $N = 256$  (right panel). The ESS was evaluated every 1k steps using 1k samples. Solid lines represent running averages of ESS across five independent training runs, with thin, light-colored lines indicating the raw ESS values from individual runs. For the local BGs, the joint efficiency is shown.  Learning rate reductions were applied after 250k and 500k steps.\label{fig:ess_comparison}}
\end{figure}

The improved training performance of the local architecture compared to a global one is further illustrated in Fig.~\ref{fig:ess_comparison}, evaluating the ESS  of both  types as a function of training time for three different systems (cubic mW ice with \(N=64\) and \(N=216\), and FCC LJ with \(N=216\)). For the global BGs, the architecture proposed in Ref.~\cite{Wirnsberger_2022} was used, which employs an attention-based coupling flow. 
Across all systems,  the local BGs consistently achieve higher ESS than the global ones at any given training time. 
This is particularly evident for the larger system sizes, demonstrating the superior scaling of the local BGs with number of particles. The local BGs also exhibit substantially lower variance in ESS, indicating that reliable estimates can be obtained even with small sample sizes and remain stable as the number of samples increases. 
\begin{table}[t]
\centering
\caption{
Effective sample size (ESS) in percent for local and global BGs for cubic mW ice and FCC LJ with different number of particles. The top three rows show the results obtained  for flow models of comparable size which were trained for 1M steps. Uncertainties are computed as the standard deviation over 4 different models each evaluated 5 times with 50k samples. The fourth column shows results reported in Ref.~\cite{Wirnsberger_2023}, which
used twice as many layers and were trained significantly longer. The results of the local BGs in the bottom six rows  were obtained using local models trained on systems with  $N = 216$ for cubic mW ice and  $N = 256$ for FCC LJ. \textbf{Bold} font indicates the best results.
}
\begin{tabular}{lcccc}
\toprule
 & Global & Local  -- joint & Local  -- marginal & Global \cite{Wirnsberger_2023} \\
\midrule
\multicolumn{5}{l}{\textit{Training}} \\
\midrule

mW ($N$=64)   & 45.9 $\pm$ 6.5  & 64.7 $\pm$ 0.7  & \textbf{82.7 $\pm$ 0.7}  & 53.7 \\
mW ($N$=216)  & 1.0  $\pm$ 0.9  & 16.4 $\pm$ 1.5  & \textbf{41.7 $\pm$ 1.6}  & 6.8 \\
LJ ($N$=256)  & 0.03 $\pm$ 0.03 & 2.0  $\pm$ 0.8  & \textbf{7.3  $\pm$ 2.6}  & - \\
\midrule
\multicolumn{5}{l}{\textit{Transfer (only local)}} \\
\midrule
mW ($N$=512)   & -  & 0.6 $\pm$ 0.4   & \textbf{1.8 $\pm$ 0.7}  & 0.2 \\
mW ($N$=1000)   & -  & 0.04 $\pm$ 0.4   & \textbf{0.1 $\pm$ 0.1}  & - \\
mW ($N$=1728)   & -  & 0.02 $\pm$ 0.02   & 0.02 $\pm$ 0.02  & - \\
LJ ($N$=500)   & -  & 0.1 $\pm$ 0.1   & \textbf{0.4 $\pm$ 0.3}  & - \\
LJ ($N$=864)   & -  & 0.03 $\pm$ 0.02   & 0.03 $\pm$ 0.03  & - \\
LJ ($N$=1372)   & -  & 0.01 $\pm$ 0.01   & \textbf{0.02 $\pm$ 0.01}  & - \\
\bottomrule
\end{tabular}
\label{tab:ess-results-transposed}
\end{table}
These trends are corroborated by the ESS obtained from the converged models, which are reported in the upper part of Table~\ref{tab:ess-results-transposed}. Remarkably, the marginal ESS of the physical system, which is the metric of primary interest, exceeds 80\% for mW ice  with $N = 64$ and over 40\% for $N = 216$, and reaches about 7\% for FCC LJ with $N=256$, highlighting the significantly improved efficiency of our local flow architecture in capturing the relevant short-range environmental information. 
Where available, we have also provided the ESS of the global BGs as reported in Ref.~\cite{Wirnsberger_2023}, which were obtained using  models comprising twice as many layers and trained significantly longer. While the final ESS reached by these models is slightly higher than the ones we obtained for the global BGs within this work, they remain far lower than those of the local BGs.

We note that the BGs do not perform equally well across all systems. For example, both local and global BGs achieve significantly higher sampling efficiencies for the cubic mW ice compared to FCC LJ, despite only a modest increase in particle number. However, we also observed that certain crystal structures within the same potential can exhibit rather different sampling efficiencies, which asks for further investigation beyond the scope of this work.

\subsection{Free energy estimation}

\begin{figure}[t]%
    \centering
    \includegraphics[width=\linewidth]{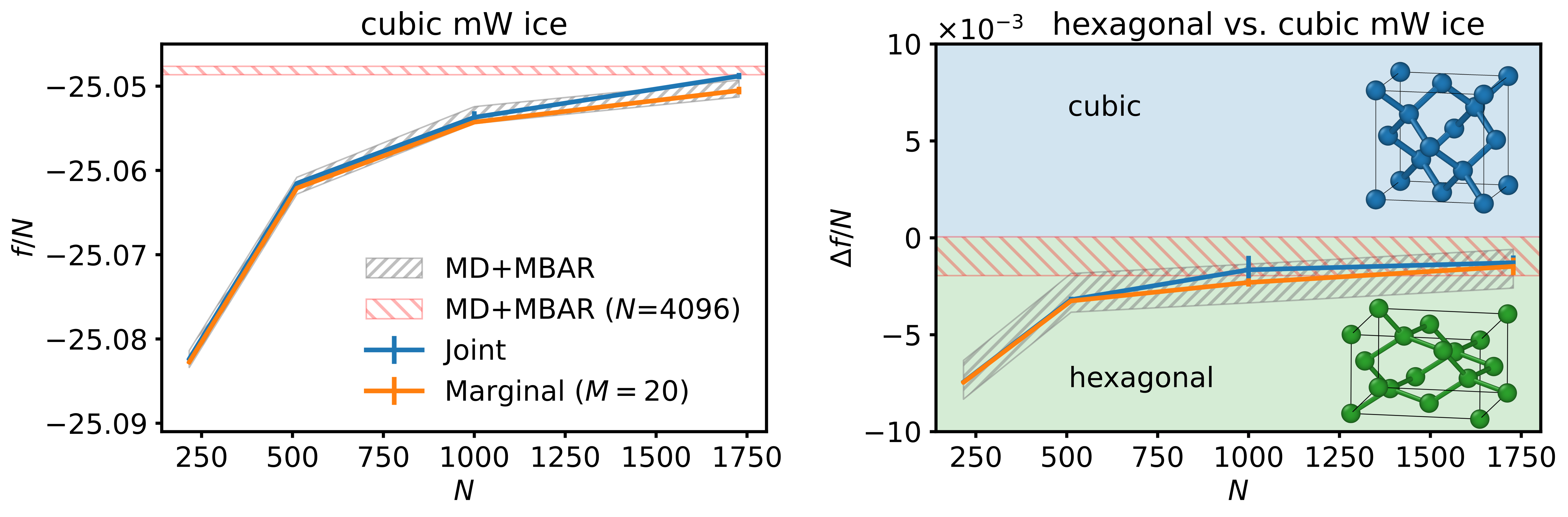}
\caption{Left: Absolute reduced Helmholtz free energy estimates per particle for the cubic  ice systems against the particle number as obtained from the local BG.  Blue and orange lines indicate the evaluation of joint and marginal densities, respectively. Reference MD+MBAR values are shown as a gray hatched area corresponding to their mean \(\pm 10^{-3}k_B T\). The red hatched are corresponds to the MD+MBAR results evaluated at $N=4096$.  BG results were obtained using a model trained $N=216$. Uncertainties were estimated by training four independent models and evaluating each of them five times with $50$k samples using different random seeds. Right: Reduced free energy difference between cubic and hexagonal ice, \(\Delta f = f_{\rm hex} - f_{\rm cubic}\), against the particle number. Hatched areas have the same meaning as for the left plot.
 \label{fig:df_transferable}}

\end{figure}
A key advantage of a trained BG is that it enables a direct evaluation of free energy differences within the framework of TFEP~\cite{tfep}, as discussed in Sec.~\ref{sec:methods}. Utilizing our local architecture, BGs trained on relatively small systems allow for an accurate evaluation of absolute Helmholtz and Gibbs free energies for far larger systems which reduces the computational effort dramatically compared to traditional free energy estimators such as MBAR~\cite{Shirts2008-mbar}.

The left panel of Fig.~\ref{fig:df_transferable} shows the absolute free energy of the cubic ice phase as a function of the number of particles computed with the BG trained on 216 particles. In addition to the joint free energy estimates, we also present values computed from the marginal generated density and further compare to high-accuracy MBAR estimates obtained from MD simulations interpolating between the Einstein crystal and the physical system of interest~\cite{Frenkel1984}. Joint and marginal free energy estimates from the BG show excellent agreement with the reference values for system sizes up to 1000 particles, deviating by less than \(10^{-3} k_B T\) per particle. 
For the largest system size ($N = 1728$), the marginal evaluation remains accurate and further improves upon the joint estimates, in agreement with the ESS results reported in Tab.~\ref{tab:ess-results-transposed}.
A similar convergence behavior  of $f$ is observed   for the FCC LJ crystal, the corresponding results are reported in the SI.

Obtaining accurate free energy difference between different crystalline phases is particularly challenging but crucial to explore phase diagrams. To reach a converged value of the free energy difference between cubic and hexagonal mW ice, system sizes up to 1728 particles are necessary, as shown in the 
right panel of Fig.~\ref{fig:df_transferable}.
Both joint and marginal estimates of $\Delta f$ are in very close agreement with the reference values, where the good performance of the joint estimates is likely due to error cancellation between the two phases.
The large number of particles necessary to fully converge the free energy difference between the cubic and hexagonal phase underscores the importance of being able to evaluate extended system size which is infeasible with global architectures.

To quantify the computational savings associated with the size-transferable free energy estimation, we compare the number of energy evaluations required to the one of the standard MD+MBAR approach using typical parameters to interpolate between the Einstein crystal and the physical system~\cite{Vega2008,Mey_Allen_Bruce,Aragones2012}. The MBAR setup requires approximately 50 intermediate states with $10^3$ uncorrelated samples per state, obtained by storing one sample every $10^3$ MD steps, resulting in a total of around $5\cdot10^7$ energy evaluations. In addition, MBAR evaluates all $5\cdot10^4$ stored samples across all 50 intermediate potentials, resulting in additional $2.5\cdot 10^6$ energy evaluations. Training the flow model until convergence involves a comparable number of evaluations, approximately $6\cdot10^7$  for $5\cdot10^5$ training steps with batches of 128 samples (see Fig.~\ref{fig:ess_comparison}). However, the training  is performed for small system sizes only. 
Once trained, evaluating the free energy with the BG requires only a few tens of thousands of samples and the corresponding energy evaluations for the large system.
In contrast, MD+MBAR necessitates several tens of millions of energy evaluations for each system size, with the computational cost scaling quadratically for pairwise potentials and even more steeply for quantum mechanical methods such as density functional theory and higher level electronic structure methods~\cite{Jones2015}. The computational savings of the flow-based approach are further amplified when performing convergence checks, which can be accomplished with a single model. Even further efficiency gains are possible by conditioning the model on external parameters, enabling reuse across multiple thermodynamic conditions~\cite{schebek2024efficient}, as exemplified below.

\subsection{Including volume fluctuations}
\begin{figure}[t]%
    \centering
    \includegraphics[width=\linewidth]{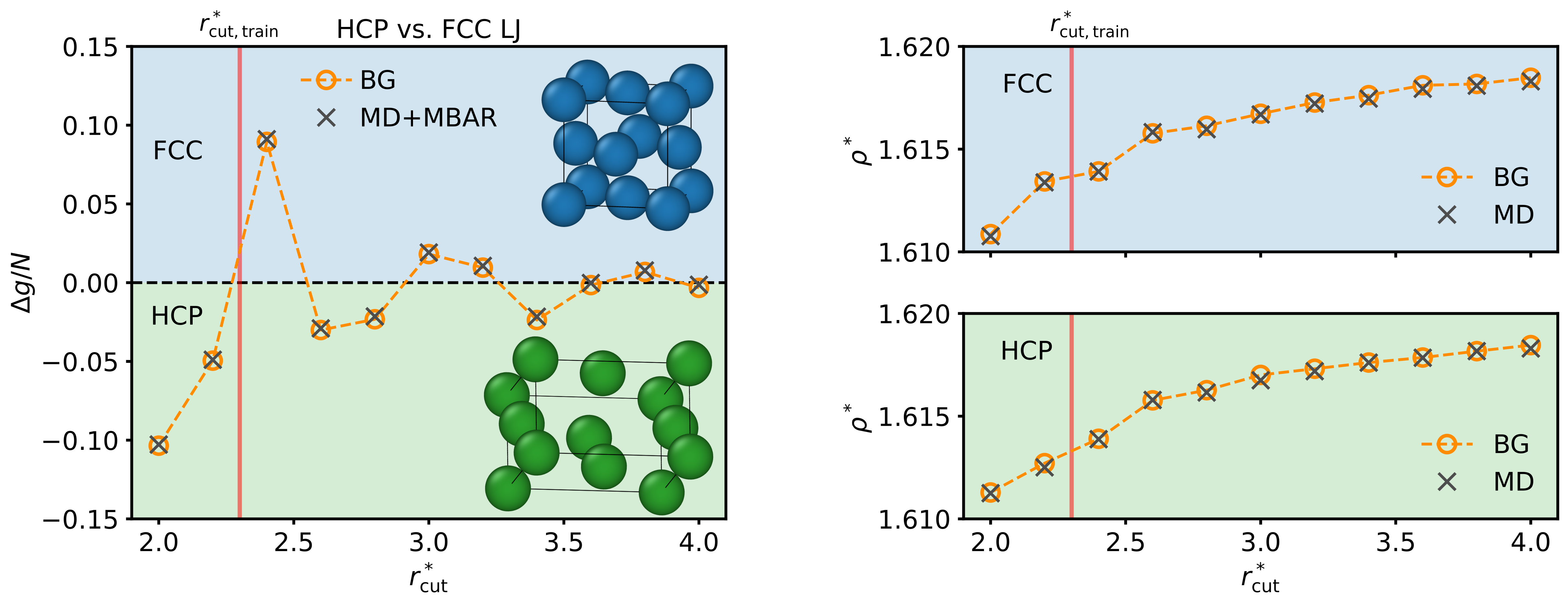}
\caption{
Left: Reduced Gibbs free energy difference \(\Delta g = g_{\rm HCP} - g_{\rm FCC}\) per particle between HCP and FCC crystal structures with 1080 particles in the LJ potential as a function of the cutoff radius. Shown are BG-based predictions (joint density estimates) alongside reference results from MD+MBAR. The red vertical line denotes the cutoff radius applied during training with 180 particles (quantities marked with an asterisk are expressed in LJ units, see SI). 
Right: Densities $\rho^*$ as obtained from the BGs (circles) and mean densities from MD $NPT$ simulations (crosses) for FCC (upper panel) and HCP (lower panel).  Error bars for both BG and MBAR are smaller than the plotted marker size (see Fig.~\ref{fig:df_transferable}).   \label{fig:dg_lj}
}
\end{figure}

To illustrate the efficiency and accuracy of our approach in the isothermal-isobaric ensemble (see Sec.~\ref{sec:scale_bg}), we determine the Gibbs free energy difference between FCC and the hexagonal close-packed (HCP) LJ crystals at constant pressure. 
Similar to the hexagonal and cubic ice phases, these structures are known to have extremely small free energy differences, which require large system sizes to converge. Crucially, the relative stability of FCC and HCP depends on the cutoff radius of the LJ potential and small cutoffs can even lead to  qualitatively wrong predictions of the stable phase ~\cite{Schieber2019, partay2027poly}.  
The local BGs for each phase are trained in a volume-conditional way (see SI) using simulation cells comprising 180 particles.   
The trained BGs are subsequently applied to systems with $N = 1080$ to evaluate the Gibbs free energy over a range of cutoff radii. 

Figure~\ref{fig:dg_lj} summarizes the BG-based Gibbs free energy estimates for the two crystalline phases.
The left panel shows the difference in reduced Gibbs free energy per particle between  HCP and FCC as a function of the cutoff radius. The  predictions of the BGs (based on joint density estimates) show excellent agreement with reference values obtained from MD combined with MBAR. Consistent with previous studies~\cite{Schieber2019, partay2027poly}, a strong dependence of the free energy difference on the cutoff radius is revealed, leading to a qualitative change in the predicted stable phase. As the cutoff increases, this sensitivity decreases, emphasizing the need for large simulation cells to accommodate sufficiently large cutoff radii. Additionally, the particle densities obtained via the Legendre transformation  using the BGs closely match the mean densities observed in  $NPT$ MD simulations, showing nearly identical dependence with respect to the cutoff radius (right panel of Fig.~\ref{fig:dg_lj}). Remarkably, 
the BG-based estimates yield highly accurate results over the entire range of cutoff radii. 
This strongly suggests that the local flow models can be trained using a single, short cutoff in a  relatively small system, yet still be transferable to much larger systems while continuing to generate accurate configurations.

\subsection{Conditioning on atom type}
\begin{figure}
    \centering
    \includegraphics[width=\linewidth]{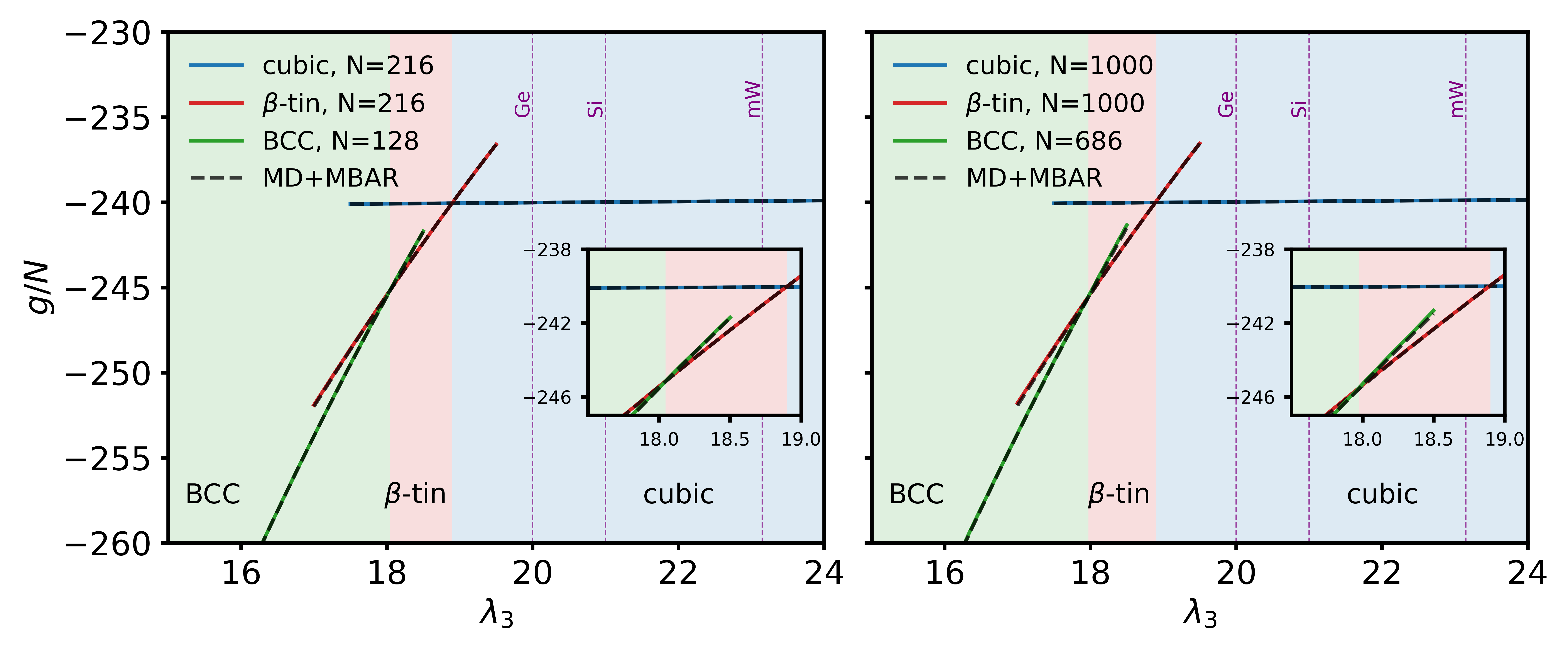}
    \caption{Gibbs free energies per particle for small (left) and large (right) particle numbers as obtained from the local BGs at zero pressure for different crystal structures as a function of the three-body interaction strength $\lambda_3$. Vertical dashed lines indicate the value of this parameter in the parametrizations for germanium, silicon, and monatomic water. The color of the shaded areas corresponds to the most stable structure. MD+MBAR values are shown as black dashed lines. Error bars for both BG and MBAR are smaller than the plotted line width.}
    \label{fig:lambda_phasediag}
\end{figure}
As another important application, we show that the flow can be trained conditioned on parameters of the interaction potential, which allows a single model to generalize over a whole class of systems. 
As an example, we employ the Stillinger-Weber (SW) potential~\cite{stillinger1985computer}, combining a pair potential
and a three-body term. The strength of the three-body interaction is controlled by a factor $\lambda_3$, yielding a total potential energy function 
%
\begin{equation}
        U_\text{SW}(\mathbf{x}) = \Phi_2(\bx) +  \lambda_3 \Phi_3(\bx) \quad.
\end{equation}
For each system of interest, the SW potential is defined in terms of characteristic length and energy scales, which are tuned to reproduce specific properties of the system. However, when expressed in reduced units, 
the only effective parameter distinguishing different parameterizations is the strength of the three-body interaction, $\lambda_3$~\cite{romano2014novel}. Importantly, different values of $\lambda_3$ stabilize or destabilize different crystal structures. 

To study the phase diagram under varying three-body strengths and system sizes, we train local BGs for diamond cubic and $\beta$-tin structures with $N=216$, and for the body-centered cubic (BCC) structure with $N=128$. The BGs are trained conditioned on $\lambda_3 \in [15, 24]$ and the shape of the simulation box, allowing the model to generalize across the full class of SW potentials and cover an entire range of different materials, as well as to evaluate Gibbs free energies. Figure~\ref{fig:lambda_phasediag} shows the Gibbs free energies  at zero pressure of the three different crystal structures as a function of $\lambda_3$ for the system sizes used during training (left panel) and transferred to $N = 1000$ for diamond cubic and $\beta$-tin and 686 for BCC (right panel).  Across the entire range of $\lambda_3 $, the  free energy estimates of the BGs align very closely with reference values from MD+MBAR. While the transition point between the cubic and $\beta$-tin structures is largely independent of system size, a slight shift is observed in the transition point between the BCC and $\beta$-tin structures, moving towards lower values of the three-body interaction strength for larger system sizes.
To capture this subtle effect, it is imperative to be able to determine highly accurate free energies for rather large system sizes. An additional advantage of the flow-based approach is that the models can be trained across all structures over the full $\lambda_3$ spectrum, even though some structures are only stable in certain parts of this region in MD simulations.

\subsection{Silicon phase diagram}
\begin{figure}
    \centering
    \includegraphics[width=.75\linewidth]{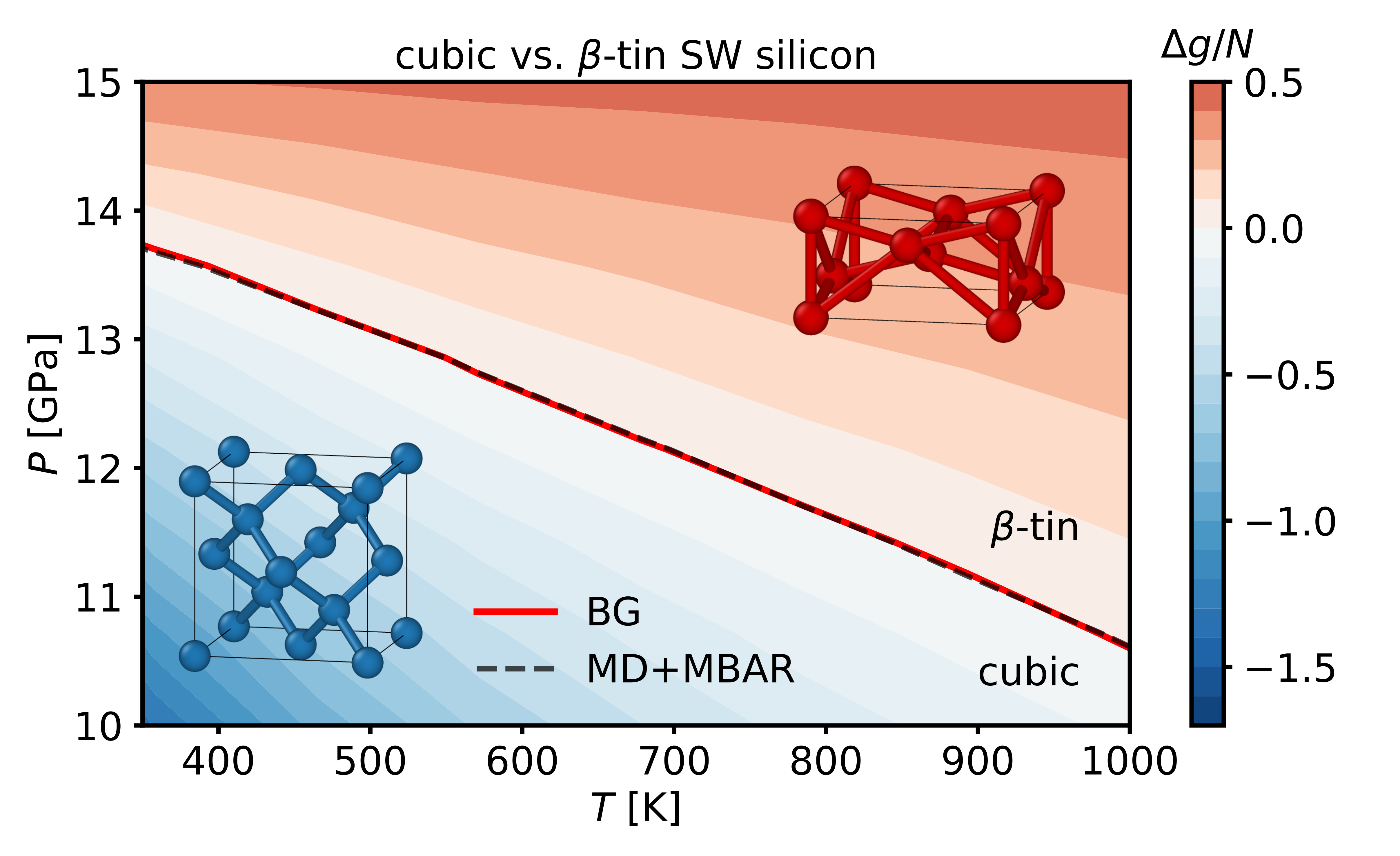}
\caption{Si phase diagram for \(N=216\) particles, computed using local BGs and MD+MBAR. The color scale represents the reduced Gibbs free energy difference per particle between $\beta$-tin and diamond cubic phases, \(\Delta g = g_{\rm cubic} - g_{\rm \beta\text{-}tin} \), as obtained from the local BGs. The red line marks the coexistence line between the two phases, while the black dashed line indicates the corresponding result from MD+MBAR estimates.  Error bars for both BG and MBAR are smaller than the plotted line width.}
    \label{fig:si_phasediag}
\end{figure}
Focusing on the SW parametrization for silicon (Si), we train local BGs conditioned on both temperature and the shape of the simulation box to map out the $(T,P)$-phase diagram. Specifically, we are aiming to determine the coexistence line between the diamond cubic and $\beta$-tin phases. Both training and evaluation of the BGs was performed with $N=216$. While size-transferability was again excellent for the cubic structure, the $\beta$-tin structure appeared to be more challenging, limiting the accuracy of free energy estimates in much larger systems.

Even without transferring to larger systems, the conditional training provides substantial cost amortization across thermodynamic states, since reference MD+MBAR calculations require fully converged simulations on at least an $7 \times 7$ grid to cover the full $(T,P)$-range, whereas the local BGs enable efficient evaluation of Gibbs free energies across all thermodynamic states. The resulting phase diagram of Si is presented in Fig.~\ref{fig:si_phasediag}. 
It captures the competition between the cubic and $\beta$-tin phases across a wide range of conditions and accurately predicts the corresponding coexistence line. Validation against free energy estimates obtained from MD+MBAR shows excellent agreement, as reflected in the practical indistinguishability of the coexistence lines.

\section{Conclusion}
We introduced a novel, local flow architecture for training generative models that enables equilibrium sampling of materials systems, scaling seamlessly from small training cells to significantly larger systems exceeding 1000 particles. Our approach surpasses previous global architectures by achieving faster training times as well as higher sampling efficiencies. Crucially, the flow models exhibit transferability across system sizes and allow to effectively sample  both short- and long-range interaction potentials, yielding highly accurate absolute free energy estimates across a variety of materials systems. By conditioning on thermodynamic variables and interaction parameters, the training process is further amortized, enabling direct and efficient evaluation of phase stability across temperature, pressure, and composition. The presented results highlight the potential of our architecture to serve as a powerful and scalable tool for studying complex materials across a wide range of thermodynamic conditions. Future directions include extending this approach to efficiently sample liquid phases and developing models with broad generalization across chemical space, thereby further expanding the scope and impact of generative modeling in materials science.

\section*{Code availability}
The flow models developed in this work will be made available at \url{https://github.com/maxschebek/bgmat}. Models are built and trained using JAX~\cite{jax2018github}, Haiku~\cite{haiku2020github} and Distrax~\cite{deepmind2020jax}. All MD simulations were performed using the OpenMM package~\cite{Eastman2023}.

\section*{Acknowledgements}
MS and JR acknowledge financial support from Deutsche Forschungsgemeinschaft (DFG) through grant CRC 1114 \ldq Scaling Cascades in Complex Systems\rdq, Project Number 235221301, Project B08 \ldq Multiscale Boltzmann Generators\rdq.  JR acknowledges financial support from DFG through the Heisenberg Programme project 428315600. MS thanks the ICC at the Flatiron Institute for hospitality while part of this research was carried out there. The computations reported in this paper were in part performed using resources made available by the Flatiron Institute. The Flatiron Institute is a division of the Simons foundation.
We thank Leon Klein, Michael Plainer, and Emil Hoffmann for insightful discussions.

 \setcounter{table}{0}
\setcounter{figure}{0}
\renewcommand{\thefigure}{S\arabic{figure}}
\renewcommand{\theequation}{S\arabic{equation}}
\renewcommand{\thetable}{S\arabic{table}}
\setcounter{equation}{0}
\setcounter{table}{0}
\setcounter{section}{0}
\renewcommand\thesection{\Alph{section}}

\newcommand{\newblock}{}
\newpage
\bibliographystyle{abbrv}
\bibliography{article}

\newpage
\section*{\large{Supplementary Information}}

\section{Augmented variables}
\subsection{Free energy}
In the following, we show that the partition function of the joint system as defined in the main text factorizes, resulting in an additive total free energy. Consider a distribution on the augmented space  of the form 
\begin{equation}\quad
    q(\bx,\ba) =\frac{ e^{-u(\bx)}\,e^{-(\mathbf{a} - \mathbf{x})^2/{2\eta^2} } }{Z^{\rm aug}}\quad.
\end{equation}
In this case, the partition function of the augmented system can be written as
\begin{align}
    Z^{\rm aug}& = \iint d\bx  d\ba \, q(\bx,\ba) = \int d\bx \,   e^{-u(\bx)}\int d\ba \, e^{-(\mathbf{a} - \mathbf{x})^2/{2\eta^2} }  \\
    & = Z \cdot Z^{\rm aux},
\end{align}
where $Z$ and $Z^{\rm aux}$ are the partition functions of  physical and auxiliary systems, respectively. In the preceding equation, we made use of the fact that the integral over the normal distribution is independent of its mean. From here, it follows immediately that
\begin{equation}
f^\text{aug} = -\log Z^\text{aug} = - \log Z-\log Z^\text{aux} = f + f^\text{aux}.
\end{equation}

\subsection{Constraining the center of mass}\label{sec:com_shift}
An important property of the potential energy in many-body systems is its invariance under global translations of the system. Therefore, optimizing the flow parameters without constraints  could lead to an uncontrolled  motion of the center of mass, distorting the free energy estimates~\cite{ahmad_free_2022, Wirnsberger_2022}. To prevent this, we follow Ref.~\citep{midgley2023eacf} and directly model the physical coordinates $\bx$ on the mean-free space $\mathbb{R}^{3N} / \mathbb{R}^{3}$, which is achieved by swapping the center of mass between physical and auxiliary systems using a ShiftCoM layer after each coupling layer~\cite{midgley2023eacf}. The ShiftCOM layer applies the  transformation
\begin{equation}
(\bx, \ba) \mapsto (\bx - \overline{\ba},\ba - \overline{\ba}),
\end{equation}
where $\overline{\ba} = \frac{1}{N} \sum_{i=1}^N \overline{\ba}_i$. Interleaving the ShiftCoM layer with the coupling layers temporarily removes the center-of-mass (CoM) constraint on the physical variables, while ensuring that the final transformation maps onto the mean-free subspace. See Ref.~\citep{midgley2023eacf} for details.

\subsection{Marginal}

The marginal of the physical system can be obtained by integrating out the auxiliary degrees of freedom, which can be approximated as~\cite{midgley2023eacf}
 \begin{equation}\label{eq:marginal}
        q_\theta(\bx') \approx  \frac{1}{M} \sum_{m=1}^M \frac{q_\theta(\bx',\ba_m)}{\pi(\ba_m|\bx')}\quad,
    \end{equation}
with $\ba_m\sim\pi(\cdot|\bx')$ and $M$ denoting the number of auxiliary samples drawn per generated physical sample. All results presented in the main text were obtained using $M=20$.
Importantly, the evaluation of the marginal distribution does only  require an inverse pass through  the flow (since $q_\theta(\bx',\ba')~=~q(f_\theta^{-1}(\bx',\ba'))|\det J_{f^{-1}_\theta}(\bx',\ba')|$), but no evaluation of the target potential.

\section{Systems}

\subsection{Lennard-Jones}

The pairwise LJ potential is given by~\cite{Frenkel2001-yl}
\begin{equation}
    u(r) = 4\epsilon \left[ {\left(  \frac{\sigma}{r} \right)}^{12}  - {\left(  \frac{\sigma}{r} \right)}^{6} \right],
\end{equation}
where $r$ is the two-particle distance. $\varepsilon$ and $\sigma$   define the characteristic length and energy scales, respectively. A cutoff radius $r_{\rm cut}$ is typically employed and the potential is shifted to be continuous at the cutoff. This results in the following interaction potential: 
\begin{equation}
    u_\text{cut}(r) = 
    \begin{cases}
        u(r)   - u(r_\text{cut})          & \text{if  }  r \leq r_\text{cut}, \\
        0                                                   & \text{else}.
\end{cases}
\end{equation}
To simplify comparisons between different systems, reduced units based on $\varepsilon$ and $\sigma$ are commonly used, and quantities in these units are typically denoted by an asterisk.

\subsection{Stillinger-Weber}
The SW potential~\cite{stillinger1985computer} features two-body ($\phi_2$) and three-body  ($\phi_3$) interactions, of which the latter enforce tetrahedral coordinations. The total potential energy is given by
\begin{equation}
    U_\text{SW}(\mathbf{x}) = \sum_i \sum_{j > i}  \phi_2(d_{ij}) +  \lambda_3 \sum_i \sum_{j \neq i} \sum_{k > j} \phi_3(d_{ij}, d_{ik}, \theta_{ijk})\quad,
\end{equation}
with
\begin{align}
  \phi_2(r) & = A \epsilon \left[ B   {\left(  \frac{\sigma}{r} \right)}^{4} - 1 \right] \exp\left(\frac{\sigma}{r - a \sigma}\right)\quad ,\\
    \phi_3(r,s, \theta) & = \lambda \varepsilon {\left( \cos\theta - \cos\theta_0  \right)}^2 \exp{ \left(\frac{\gamma \sigma}{r - a \sigma}\right)} \exp{\left(\frac{\gamma \sigma}{s - a \sigma}\right)},
\end{align}
where $d_{ij}$ denotes a two-particle distance and $\theta_{ijk}$ is the angle formed by the three atoms. All parameters are fixed except for $\varepsilon$, $\sigma$, and $\lambda_3$, which are adjusted to model a specific system. Here, $\lambda_3$ sets the strength of the three-body interactions, while $\varepsilon$ and $\sigma$ define the characteristic energy and length scales. The common constants are $A = 7.049556277$, $B = 0.6022245584$, $a = 1.8$, $\theta = 109.47^\circ$, and $\gamma = 1.2$. For the monatomic water (mW) model, the tuned values are $\lambda = 23.15$, $\varepsilon = 6.189~\text{kcal/mol}$, and $\sigma = 2.3925~\text{\AA}$~\cite{molinero2009water}, while for silicon $\lambda = 21$, $\varepsilon = 50.003~\text{kcal/mol}$, and $\sigma = 2.0951~\text{\AA}.$ As in the Lennard–Jones potential, reduced units can be adopted in which $\varepsilon$ and $\sigma$ serve as the energy and length scales. In this representation, the only parameter that differentiates models is $\lambda_3$.

\subsection{Computational settings}

\textbf{NVT simulations}\\
\noindent Cubic and hexagonal mW ice was modeled at \( T = 200K \) at a density of $\rho=1.004$ gcm$^{-3}$. The FCC LJ crystal was modeled at $T^*=2.0$ and $\rho^*=1.28$.\\

\noindent  \textbf{NPT simulations}\\
\noindent The FCC and HCP LJ crystals were simulated at a temperature of $T^*=0.2$ and the Gibbs free energy was evaluated at $P^*=150$. The phase diagram of the SW potential was simulated at a reduced temperature of $T^*=0.01$. 

\subsection{Unit cells}
Table~\ref{tab:unitcells} summarizes the unit cells used for the different crystal structures.

\begin{table}[h!]
\centering
\caption{Unit cell shapes and number of atoms for different crystal structures.}
\begin{tabular}{lccc}
\toprule
Crystal Structure & Unit Cell Shape & Number of Atoms per Unit Cell \\
\hline
cubic ice           & cubic        & 8  \\
diamond cubic          & cubic        & 8  \\
hexagonal ice  & orthorhombic        & 8 \\
FCC & (cubic / orthorhombic)        & (4 / 6)  \\
HCP                     & orthorhombic        & 4  \\
BCC               & cubic & 2 \\
$\beta$-tin               & orthorhombic & 4 \\
\bottomrule
\end{tabular}
\label{tab:unitcells}
\end{table}

\section{Model details}
\subsection{Graph neural network}
The graph neural network used for computing the particle embeddings is composed of $L$ layers, where the update of the $l$th layer is calculated as
\begin{equation}\label{eq:gnn}
\begin{aligned}
\bd_{ij} &= {\rm sinusoidal}([ \ba_i - \ba_j ]_{\rm PBC}; \boldsymbol{\omega}_d)\quad, \\[5pt]
\bm^l_{ij} &= \phi_e(\bh_i^l,\bh_j^l,\mathbf{d}_{ij})\quad, \\[5pt]
\bm^l_i &= \sum_{j\in\mathcal{N}_i} \bm^l_{ij}\quad, \\[5pt]
\bh_i^{\ba,l+1} &= \phi_h(\bh_i^{\ba,l},\bm^l_i)\quad.
\end{aligned}
\end{equation}
$\mathcal{N}_i$ contains all particles defined as the local neighborhood around the auxiliary particle $i$ and $\phi_e$ and $\phi_h$ are implemented as multilayer perceptrons (MLP). The notation $ [\cdot]_{\rm PBC}$ indicates that the difference vector is to be computed respecting periodic boundary conditions. To increase expressivity, we make use of sinusoidal embeddings (see following subsection) for both the difference vectors and the initial node embedding, using separate base frequencies $\boldsymbol{\omega}\in\mathbb{R}^3$ for each. 

\subsection{Sinusoidal encoding}
We use sinusoidal encodings to encode difference vectors and ideal lattice positions. Initial embeddings are set as $\mathbf{h}_i^0 = {\rm sinusoidal}(\mathbf{x}_i^0; \boldsymbol{\omega}_{\rm init})$, where $\mathbf{x}_i^0$ is the ideal lattice positions of atom $i$. We set $\boldsymbol{\omega}_{\rm init}$ to match the unit cell, so atoms in repeating cells share embeddings. Longer frequencies that uniquely label each particle perform similarly, but the unit-cell-based approach is naturally transferable to any supercell.
For the sinusoidal encoding of dimension $j$ of a vector $\mathbf{x}\in\mathbb{R}^3$, we use~\cite{Wirnsberger_2022}
\begin{equation}
\begin{aligned}
    {\rm sinusoidal}(\bx,\boldsymbol{\omega})_j =  [&\cos(\omega_j x_{j})\quad,
    \sin(\omega_j x_{j}),\\
    &\cos(2\omega_j x_{j}), \sin(2\omega_j x_{j})\quad,\\
    &\ldots,\\
    &\cos(N_{f}\omega_j x_{j}), \sin(N_f\omega_j x_{j}]\quad,
\end{aligned}
\end{equation}
where $N_f$ is the total number of frequencies and $\boldsymbol{\omega}\in\mathbb{R}^3$ contains the base frequencies for the three spatial dimensions.

\subsection{Conditional training}

Conditioning on external parameters such as the shape of the simulation box, parameters of the interaction potential, or thermodynamic states, is achieved by minimizing the conditional loss function
\begin{equation}\label{eq:cond_loss}
    \mathcal{L}_{qp}(\theta)  = - \mathbb{E}_{\bh\sim p^{\phantom{T}}_{\bc}} \mathbb{E}_{\bx\sim q}\bigl[\log w(\bx|\bc)\bigr]\quad,
\end{equation}
where \(p^{\phantom{T}}_{\bc}\) defines the distribution of the conditional parameter $\bc$. If no prior information about the system’s shape distribution is available, a reasonable estimate can be obtained from short MD simulations at very small system sizes.
\newpage

\subsection{Hyperparameters}
\begin{table}[ht]
\caption{\label{tab:model_hyper}Model hyperparameters of the local augmented flow.}
\begin{tabularx}{\textwidth}{XX} 
\toprule
\textbf{Normalizing flow} & \\ 
Number of layers & $10$ (mW/SW), $8$ (LJ) \\
\\
\textbf{Graph neural network} & \\ 
Number of layers & $2$ \\
Embedding dimension & $96$ \\
Number of frequencies in sinusoidal embedding ($N_f$) & 
$8$\\
Number of neighbors for message passing ($\mathcal{N}_i$) & $16$ (cubic ice, diamond cubic), $17$ (hexagonal ice), $12$ (HCP and FCC LJ, 14 (BCC), 16 ($\beta$-tin)\\ 

\\
\textbf{Circular rational-quadratic spline} & \\ 
Number of segments & $16$\\
\\
\textbf{Base distribution} & \\ 
$\eta_q$
&
$0.05\sigma$ (LJ), $0.2$ \AA (mW/SW)
\\
\bottomrule
\end{tabularx}
\end{table}

\begin{table}[ht]
\caption{\label{tab:model_hyper}Model hyperparameters of the global flow.}
\begin{tabularx}{\textwidth}{XX} 
\toprule
\textbf{Normalizing flow} & \\ 
Number of layers & $12$ \\
\\
\textbf{Transformer} & \\ 
Number of blocks & $2$ \\
Number of heads &  $2$ \\
Embedding dimension & $128$ \\
Number of frequencies in sinusoidal encoding ($N_f$) & 
$8$ (mW $64$ particles), $16$ (mW $216$ and LJ $256$ particles)\\
\\
\textbf{Circular rational-quadratic spline} & \\ 
Number of segments & $16$\\
\\
\textbf{Base distribution} & \\ 
$\eta_q$
&
$0.01\sigma$ (LJ), $0.2$ \AA (mW, SW)
\\
\bottomrule
\end{tabularx}
\end{table}

\section{Flow-based Gibbs free energy calculations}

For the minimization $G = \min_\bh[F(\bh) + \det(\bh) P]$, $F(\bh)$ is evaluated using $10^3$ samples, while $10^4$ samples were used for a final evaluation of $G$ at the optimal box parameteres. To reduce memory costs, $\tilde{G}(\bh)$ is minimized using gradient-free methods like Nelder--Mead~\cite{Nelder1965-dn}, typically converging within a few tens of steps. Only isotropic deformations were allowed, except for $\beta$-tin, where the box was flexible in all three dimensions. For MD-based calculations, $\langle \bh \rangle_{T,P}$ is obtained from $NPT$ simulations and $F(\langle \bh \rangle)$ from $NVT$, giving $G = F + P\det \langle \bh \rangle_{T,P}$. The two approaches reproduce the same result when the minimizing box matches the mean observed in MD, which holds for all systems investigated and is consistent with the harmonic approximation.

\section{Free energy estimates}
\subsection{Helmholtz Free energies}

\begin{figure}[h]%
    \centering
    \includegraphics[width=\linewidth]{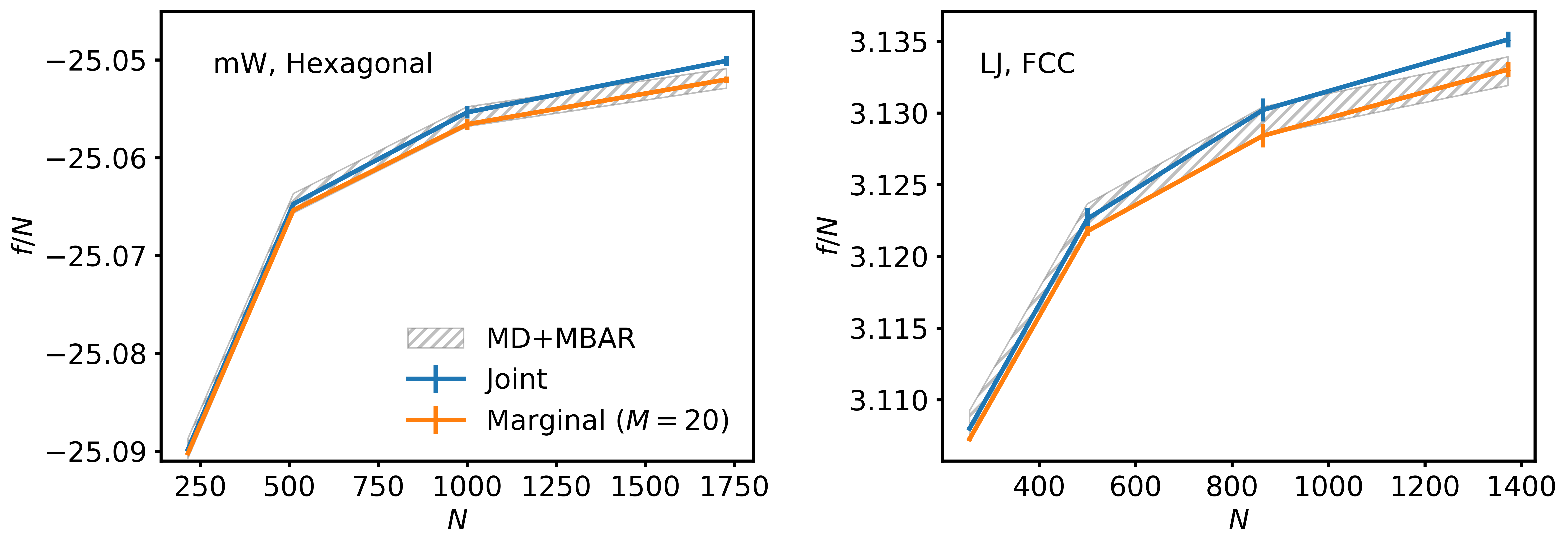}
\caption{Reduced absolute Helmholtz free energies of  the hexagonal ice in the mW potential (left) and the FCC crystal in the LJ potential against the number of particles.}

\end{figure}


\section{Training costs}
Unless stated otherwise, all models were trained on NVIDIA A5000 GPUs, using four GPUs in parallel. For the mW ice system with \( N = 216 \), training the local flows took approximately four days for 1 million steps, while the global flows required around 2 days. For the LJ system, the local flows took about 3 days, and the global flows took around 2 days. The smaller mW system with \( N = 64 \) required around 2 days of training on a single GPU for the global flow and and around 3 days  for the local flow. To enable a direct comparison of training costs with the results of Ref.~\cite{Wirnsberger_2022,Wirnsberger_2023}, we also trained one model on cubic mW ice with \( N = 216 \) using four NVIDIA A100 GPUs. The training required two days for one million steps, with convergence reached after 500k steps (or approximately one day).

\section{Optimization details}

We used the Adam optimizer~\cite{kingma_adam} for the training of all models and used gradient clipping for stability. Further, the initial learning rate of $10^{-4}$ was reduced after 250k and 500k steps by a factor of 10. We followed Ref.~\cite{Wirnsberger_2022} and trained on a linearized version of the interaction potentials, defined by
\begin{equation}
    u_{\mathrm{lin}}(r) = \begin{cases} 
    u(r_{\mathrm{lin}}) + u'(r_{\mathrm{lin}})(r - r_{\mathrm{lin}}) & r < r_{\mathrm{lin}} \\
u(r) & r\ge r_{\mathrm{lin}}, \end{cases}
\end{equation}
where $u$ is the original pairwise potential and $r_{\mathrm{lin}}$ is a distance threshold below which it is linearized. Setting $r_{\mathrm{lin}}$ smaller than the typical distance between particles stabilizes training without biasing the target distribution.

\end{document}